\documentclass{article}

\usepackage{microtype}
\usepackage{graphicx}
\usepackage{booktabs} %
\usepackage{xcolor}
\usepackage{hyperref}

\usepackage[font=small]{caption}

\usepackage[accepted]{mlsys2022}

\usepackage{amsmath}  %

\usepackage[frozencache,cachedir=.]{minted}  %

\usepackage{caption}
\usepackage{subcaption}  %
\usepackage{xspace}  %
\usepackage{xurl} %
\usepackage{float}
\usepackage{enumitem} %

\newcommand{\IGNORE}[1]{}

\newcommand{\ourname}[0]{\textsc{Pathways}\xspace}
\newcommand{\plaquename}[0]{\textsc{Plaque}\xspace}
\newcommand{\google}[0]{Google\xspace}  %

\newcommand{\sectionshrinker}{}
\newcommand{\subsectionshrinker}{}
\newcommand{\captionshrink}{}
\newcommand{\extrashrink}[1]{}

{\end{list}}

{\end{list}}

\mlsystitlerunning{Pathways: Asynchronous Distributed Dataflow for ML}

\begin{document}

\twocolumn[
\mlsystitle{Pathways: Asynchronous Distributed Dataflow for ML}

\begin{mlsysauthorlist}
\mlsysauthor{Paul Barham}{g}
\mlsysauthor{Aakanksha Chowdhery}{g}
\mlsysauthor{Jeff Dean}{g}
\mlsysauthor{Sanjay Ghemawat}{g}
\mlsysauthor{Steven Hand}{g}
\mlsysauthor{Dan Hurt}{g}
\mlsysauthor{Michael Isard}{g}
\mlsysauthor{Hyeontaek Lim}{g}
\mlsysauthor{Ruoming Pang}{g}
\mlsysauthor{Sudip Roy}{g}
\mlsysauthor{Brennan Saeta}{g}
\mlsysauthor{Parker Schuh}{g}
\mlsysauthor{Ryan Sepassi}{g}
\mlsysauthor{Laurent El Shafey}{g}
\mlsysauthor{Chandramohan A. Thekkath}{g}
\mlsysauthor{Yonghui Wu}{g}
\end{mlsysauthorlist}

\mlsysaffiliation{g}{Google}

\mlsyscorrespondingauthor{\ourname authors}{pathways-mlsys@google.com}

\mlsyskeywords{machine learning, dataflow, accelerators}

\vskip 0.3in

\begin{abstract}

We present the design of a new large scale orchestration layer for accelerators. Our system,~\ourname, is explicitly designed to enable exploration of new systems and ML research ideas, while retaining state of the art performance for current models. \ourname uses a \emph{sharded} dataflow graph of \emph{asynchronous} operators that consume and produce futures, and efficiently gang-schedules \emph{heterogeneous} parallel computations on thousands of accelerators while coordinating data transfers over their dedicated interconnects. \ourname makes use of a novel \emph{asynchronous distributed dataflow} design that lets the control plane execute in parallel despite dependencies in the data plane. This design, with careful engineering, allows~\ourname to adopt a single-controller model that makes it easier to express complex new parallelism patterns. 
We demonstrate that \ourname can achieve performance parity ($\sim100\%$ accelerator utilization) with state-of-the-art systems when running SPMD computations over 2048 TPUs, while also delivering throughput comparable to the SPMD case for Transformer models that are pipelined across $16$ stages, or sharded across two islands of accelerators connected over a data center network.

\end{abstract}

]

\printAffiliationsAndNotice{}  %

\widowpenalty0
\clubpenalty0

\sectionshrinker
\section{Introduction}\label{sec:intro}

Deep learning has seen remarkable achievements over the last
decade, across domains from image
understanding~\cite{krizhevsky2012imagenet,he2016deep} to natural language
processing~\cite{devlin2019bert,brown2020language}. This rapid recent progress of machine
learning (ML) has been characterized by the co-evolution of ML models,
accelerator hardware, and the software systems that tie the two
together. This co-evolution poses a danger that 
systems become over-specialized to \emph{current} workloads and fail to
anticipate future needs. In this paper, we describe \ourname, a new system built for distributed ML. \ourname is designed to target
specific capabilities that we believe will be needed by future ML workloads~\cite{2021pathwaysarchitecture} -- and are therefore needed \emph{today} to support research into those workloads -- but which are poorly supported by state-of-the-art systems.

For example, most of today's state-of-the-art ML workloads use a ``single program multiple data'' (SPMD) model, inspired by MPI~\cite{clarke1994mpi}, where all accelerators run the same computation in lockstep and communication between accelerators is described by collectives like AllReduce. Recently, researchers have begun to run into the limits of SPMD for ML computations. Very large language models have been scaled up using pipelining rather than pure
data-parallelism~\cite{narayanan2019pipedream,rasley2020deepspeed, narayanan2021efficient}, and models such as Mixture of Experts (MoE)~\cite{shazeer2017outrageously} have started to explore computational sparsity that is most naturally expressed using
fine-grain control flow and heterogeneous computation across
accelerators. System designers have adopted ingenious techniques to
execute pipelined~\cite{narayanan2021efficient,rasley2020deepspeed,narayanan2019pipedream, huang2019gpipe} and
\emph{homogeneous} MoE~\cite{lepikhin2020gshard, switchtransformer2021} models on MPI-style
systems, but as we argue in detail later, the MPI programming model is too restrictive both for users and for the underlying system.

On the other hand, with each new generation of accelerators, ML clusters are becoming increasingly heterogeneous~\cite{jeon2019philly,chaudhary2020gandivafair,weng2022mlaas}. Providing exclusive access to large ``islands'' of homogeneous accelerators connected over high-bandwidth interconnects is expensive, and often wasteful as a single user program must try to keep all of the accelerators continuously busy. Such constraints are further driving researchers towards ``multiple program multiple data'' (MPMD) computations that allow more flexibility by mapping sub-parts of the overall computation to a collection of more readily available smaller islands of accelerators. To increase utilization, some ML hardware resource management researchers~\cite{xiao2020antman,bai2020pipeswitch,yu2020fine,wang2021ticktock,lim2021zico,zhao2022vpipe,weng2022mlaas} multiplex hardware in a fine-grained manner between workloads, enabling workload elasticity, and improving fault tolerance.

Finally, researchers are beginning to standardize on a set of \emph{foundation models}~\cite{bommasani2021foundation, 2021pathwaysarchitecture} that are trained on large data at scale and are adaptable to multiple downstream tasks. Training and inference for such models offers opportunities for improving cluster utilization by multiplexing resources across many tasks, and efficiently \emph{sharing} state between them.  For example, several researchers might concurrently fine-tune~\cite{houlsby2019parameter, zhang2021share} a foundation model for different tasks, using the same accelerators to hold the fixed foundation model layers. Training or inference over shared sub-models can benefit from techniques that allow examples from different tasks to be combined in a single vectorized batch to get better accelerator utilization~\cite{crankshaw2017clipper}.

This paper describes our system, \ourname, which matches the functionality and performance of state of the art ML systems, while providing the capabilities needed to support future ML workloads. \ourname uses a client-server architecture that enables \ourname's runtime to execute programs on system-managed islands of compute on behalf of many clients.  \ourname is the first system designed to transparently and efficiently execute programs spanning multiple ``pods'' of TPUs~\cite{googlecloudtpu}, and it scales to thousands of accelerators by adopting a new dataflow execution model.
\ourname's programming model makes it easy to express non-SPMD computations and enables centralized resource management and virtualization to improve accelerator utilization.

In the remainder of the paper we first discuss the limitations of current distributed ML systems and motivate our design choices for \ourname (\S\ref{sec:singlevsmulti}), and next describe the flexible programming model that \ourname supports (\S\ref{sec:impl:programmingmodel}). We describe \ourname's architecture (\S\ref{sec:backend}), highlighting how we have addressed the key limitations
of older client-server ML systems using a \emph{sharded dataflow model} and \emph{asynchronous gang-scheduling}.
We present both micro-benchmarks and end-to-end evaluations using real ML models  that demonstrate we have met the goal of matching the performance of state-of-the-art multi-controllers for realistic workloads (\S\ref{sec:evaluation}), and
validate that \ourname's mechanisms are well suited to support the features 
needed for the research and deployment of novel and efficient ML methods.

\section{Design Motivation}\label{sec:singlevsmulti}

\begin{figure*}
\centering
\begin{subfigure}[t]{0.26\linewidth}
\raisebox{10pt}{
\includegraphics[scale=0.49]{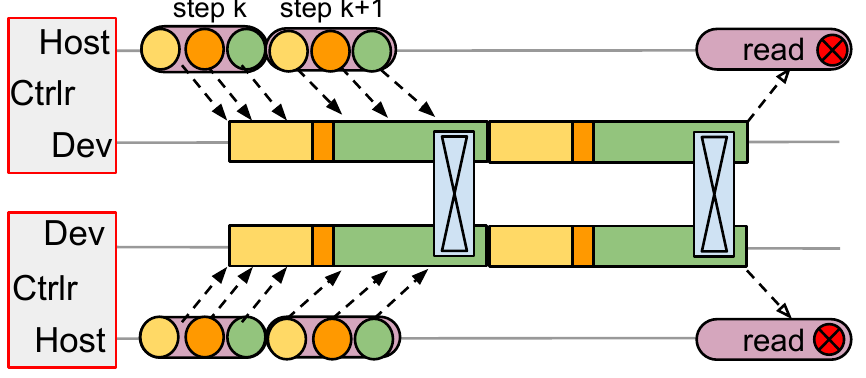}
}
\subcaption{\small JAX/PyTorch SPMD}
\label{subfig:mc_spmd}
\end{subfigure}
\begin{subfigure}[t]{0.33\linewidth}
\raisebox{8pt}{
\includegraphics[scale=0.47]{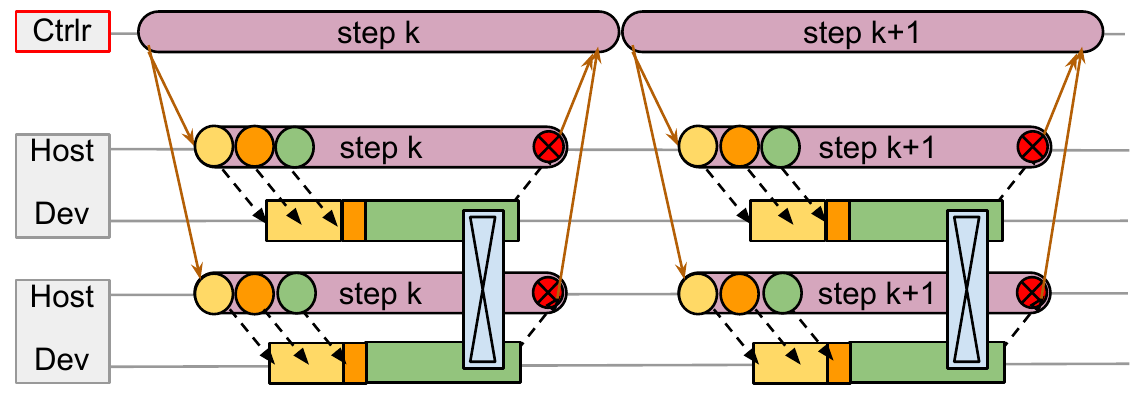}}
\captionshrink
\subcaption{TF1 SPMD}
\label{subfig:sc_spmd}
\end{subfigure}
\begin{subfigure}[t]{0.26\textwidth}
\includegraphics[scale=0.39]{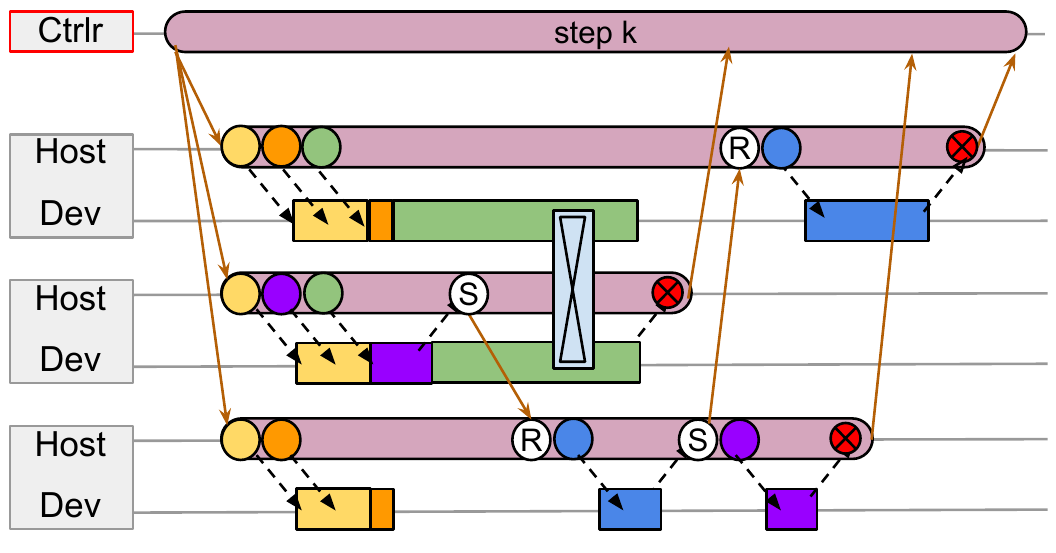}
\subcaption{\small TF1 non-SPMD}
\label{subfig:sc_non_spmd}
\end{subfigure}
\begin{subfigure}[t]{0.12\textwidth}
\raisebox{-20pt}{
\includegraphics[scale=0.5]{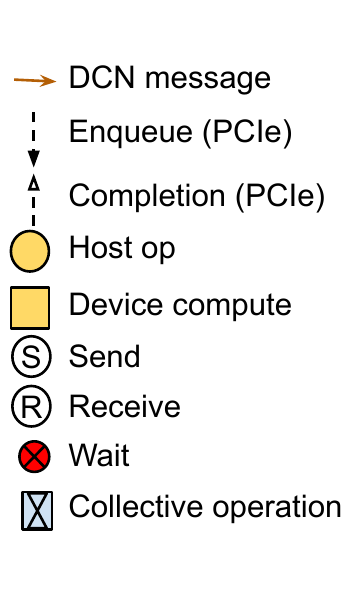}}
\end{subfigure}
\captionshrink
\caption{Comparison of dispatch overheads and communication patterns between multi-controller and single-controller systems.  (a) Jax or PyTorch SPMD independently enqueues accelerator computations asynchronously over fast PCIe; (b) TensorFlow v1\@ SPMD requires control messages over slower DCN; (c) TensorFlow v1\@ non-SPMD programs require cross-host coordination or data transfer through explicit send~(S) 
and recv~(R) ops. 
}
\label{fig:single-multi-controller}
\vspace{-2ex}
\end{figure*}

The design choices of distributed ML systems are often driven by the properties of the underlying target hardware accelerators. We refer readers to Appendix~\ref{sec:background} for a discussion on some of these properties  and how they typically influence distributed ML systems. Here, we focus on how some of the design and implementation choices of existing distributed ML systems make it hard for them to support large, sparse or irregular models. 

Distributed ML systems for training state-of-the-art SPMD models often adopt a \emph{multi-controller} architecture where the same client executable is run directly on all the hosts in the system, taking exclusive ownership of the resources on those hosts for the duration of the program execution. Examples of this architecture include MPI~\cite{clarke1994mpi}, PyTorch~\cite{paszke2019pytorch}, JAX~\cite{jax2018github}, and more recent configurations of TensorFlow~\cite{shazeer2018mesh, agrawal2019tensorflow}. The key advantage of this architecture is the low latency for dispatching accelerator computations (see Figure~\ref{subfig:mc_spmd}) since an identical copy of the user's code runs on each of the accelerator hosts and dispatch involves communication only over (relatively) fast PCIe links. All other communication across hosts only happens through collectives that use dedicated interconnects like NVLink~\cite{foley2017ultra} and ICI~\cite{jouppi2020tpu} without going via host memory. However, this architecture is a poor match for modern ML workloads that use pipelining or computational sparsity. Any communication beyond standard collectives in multi-controller systems requires users to implement their own coordination primitives.  The multi-controller approach also typically assumes exclusive ownership of hardware resources. This not only shifts the responsibility of ensuring high utilization of the expensive accelerators on to the user, but also complicates the design of features like resource virtualization and multiplexing that are needed to build efficient cluster-wide ML infrastructure.

Single-controller systems such as TensorFlow v1~\cite{abadi2016tensorflow} offer a very general distributed dataflow model, including
optimized in-graph control flow~\cite{yu2018dynamic}. 
A TensorFlow~(TF) Python client builds a
computation graph and hands it off to a coordinator runtime, which
partitions the graph into a subgraph for each worker and delegates the
execution of the subgraphs to local runtimes on workers.
Coordination between workers is performed using data- and control-edges passing messages over the datacenter network (DCN). While the single-controller design offers a flexible programming model and virtualization of resources, it presents implementation challenges. 

Firstly, while multi-controller systems only require communication over PCIe to dispatch accelerator computations (Figure~\ref{subfig:mc_spmd}), clients in single-controller systems are ``farther away" and the dispatch latency involves communication over DCN, typically an order of magnitude slower than PCIe (Figure~\ref{subfig:sc_spmd}). Secondly, to support concurrent execution of MPMD programs with SPMD sub-computations, each spanning a subset of accelerators drawn from a shared cluster, the runtime must have some mechanism to support gang-scheduling of accelerator computations. Gang-scheduling is essential in the case of TPUs, since they are single-threaded and only run non-preemptible kernels, so the system will deadlock if communicating computations are not enqueued in a consistent order. Even for GPUs or other accelerators that can execute concurrent computations, gang scheduling allows more efficient execution of collectives~\cite{feitelson1992gs}. Single-controller systems for ML therefore require a distributed scheduling mechanism to order the computations enqueued on behalf of different programs. Finally, a system for modern ML workloads must be designed to run computations distributed over thousands of accelerators, with first class support for sharded representations and data structures. For instance, a naive dataflow graph representing an edge between an M-way sharded computation and an N-way sharded computation would require $M+N$ nodes and $M\times N $ edges, rapidly becoming unwieldy.

The implementation choices made by TF v1 were over-specialized to assume a single, smallish, exclusively-owned island of accelerators. This over-specialization makes it practically infeasible to use TF for contemporary or future ML workloads. While TF can run computations that require cross-host coordination or data transfer through send and recv ops (Figure~\ref{subfig:sc_non_spmd}),  host side work at the destination like dispatching the accelerator computation is triggered only after the transfer is completed. In programs involving many cross-host transfers, for example pipelined models with a large number of stages, these dispatch latencies accumulate, leading to inefficient accelerator utilization. %
 While TF v1 users can (inefficiently) enforce a consistent ordering for gang-scheduling within a single program, by using control edges, the lack of a centralized scheduler in single-controller systems like TF v1 makes it impossible to ensure consistent ordering between computations \emph{across} programs. TF also materializes the full sharded computation graph, which introduces substantial overhead in both graph serialization and execution when the number of shards reaches into the thousands, leading to millions of graph edges between sub-computations.

\ourname combines the flexibility of single-controller frameworks with the performance of multi-controllers. We adopt a single-controller model since we believe it offers much better opportunities than multi-controller for \emph{novel} and \emph{efficient} ML computation, both by exploiting computational sparsity and heterogeneity, and by enabling cluster management systems that promote sharing and virtualizing resources. Our design differs from older single-controller ML systems in that it uses asynchronous dispatch to match the performance of multi-controller systems, supports centralized resource management and scheduling with first-class support for gangs of SPMD accelerator computations, and uses a sharded dataflow system for efficient coordination.

\section{\ourname Programming Model}\label{sec:impl:programmingmodel}

We have implemented support to target \ourname from source programs written in TensorFlow and JAX, but we concentrate on JAX for the evaluation in this paper. JAX users can explicitly wrap standard Python code
with decorators to indicate fragments that should be compiled into (potentially SPMD) XLA computations. These XLA computations are usually characterized by known input and output types and shapes, bounded loops, and with few (if any) conditionals (see Appendix~\ref{sec:mlprogram} for more details) making it feasible to estimate the resource requirements of computations in advance.  We refer to these computations with known resource requirements as ``compiled functions". Each such function maps to a single (sharded) computation node in a \ourname program.

JAX today cannot scale beyond a single TPU pod since JAX programs that run in multi-controller configurations transfer all data using XLA collectives, and 
these are only currently available over ICI on TPU. 
\ourname can be used as a plug-in replacement for the %
JAX backend, allowing JAX code to run unmodified except that SPMD computations now have access not just to the locally connected TPU cores, but to as many cores as are provisioned in the system.
And since \ourname can communicate over both ICI and DCN, it allows JAX programs to scale for the first time to multiple TPU pods, containing many thousands of TPU cores.

The ability to run unmodified JAX code is convenient but does not unlock the full performance of \ourname.  A \ourname user may request sets of ``virtual devices'', with optional constraints on the device types, locations or interconnect topology, and is then able to place specific compiled functions on those devices (Figure~\ref{fig:user-code}).  The system will automatically handle all data movement and resharding between dependent computations.

By default, we convert each compiled function into a standalone \ourname program containing just one (sharded) computation, meaning that if a user wants to run many functions back to back, a separate Python call and RPC from client to coordinator is required for each function. We therefore also implemented a new \emph{program tracer}~(Figure~\ref{fig:user-code}) that a user can wrap around a block of Python code that calls many compiled functions. The tracer generates a single \ourname program where each compiled function is represented by a computation node in a dataflow graph.

\begin{figure}[t]
\vspace{0.8ex}
\centering
\begin{minted}[fontsize=\scriptsize,frame=lines]{python}
def get_devices(n):
  """Allocates `n` virtual TPU devices on an island."""
  device_set = pw.make_virtual_device_set()
  return device_set.add_slice(tpu_devices=n).tpus

a = jax.pmap(lambda x: x * 2., devices=get_devices(2))
b = jax.pmap(lambda x: x + 1., devices=get_devices(2))
c = jax.pmap(lambda x: x / 2., devices=get_devices(2))

@pw.program  # Program tracing (optional)
def f(v):
  x = a(v)
  y = b(x)
  z = a(c(x))
  return (y, z)

print(f(numpy.array([1., 2.])))
# output: (array([3., 5.]), array([2., 4.]))
\end{minted}
\vspace{-2.5ex}
\captionshrink
\caption{Python user code example for \ourname running sharded computations across multiple islands of TPU.}
\label{fig:user-code}
\vspace{-2ex}
\end{figure}

JAX's philosophy of supporting \emph{transforms} of traced code is a good match for the research directions we want to explore. For example, JAX has a companion library called FLAX~\cite{flax2020github} that is used to express layered DNN models, and we have written a library that automatically converts a FLAX model into a pipelined \ourname program. In addition, JAX supports transforms to vectorize ``per-example'' Python functions, producing efficient batched code, and such transforms are a good basis for exploring new forms of data-dependent vectorized control flow, as we briefly describe later (\S\ref{sec:discussion-controlflow}).

\section{\ourname System Architecture}\label{sec:backend}

\begin{figure*}
\includegraphics[height=1.7in]{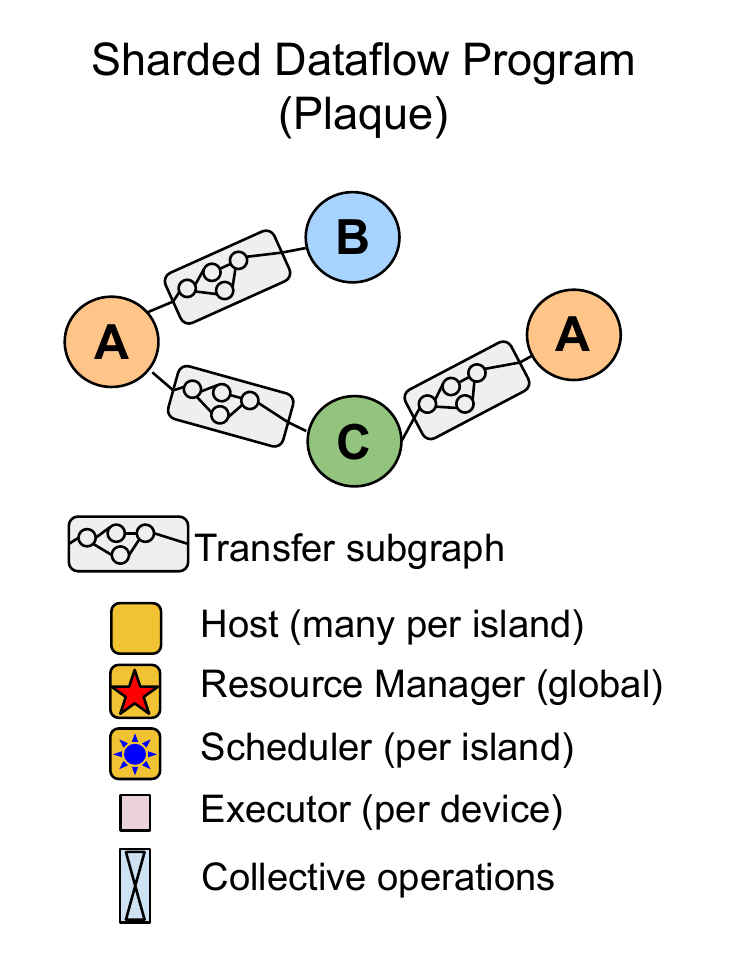}
\hspace{4mm}
\includegraphics[height=1.7in]{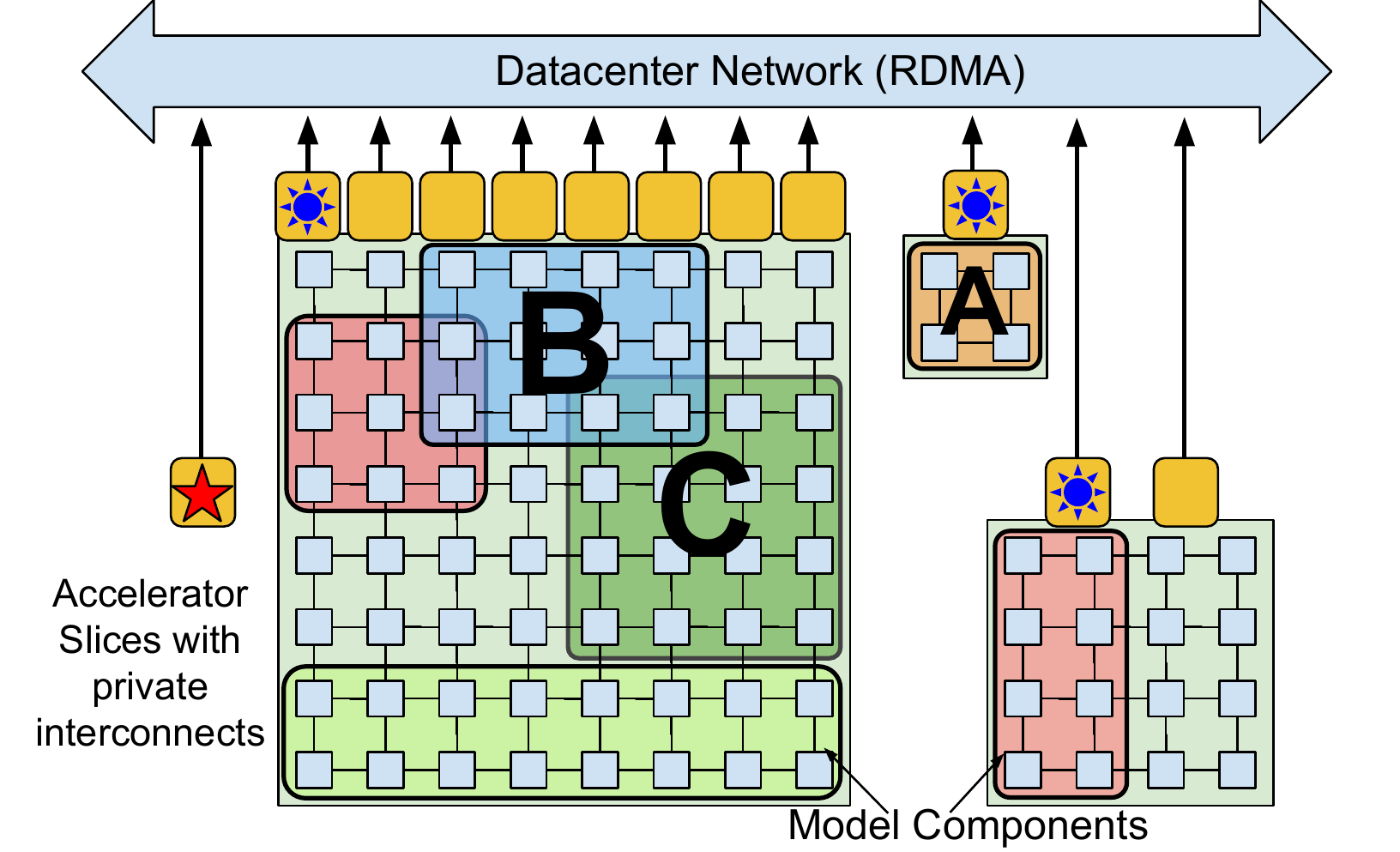}
\hspace{4mm}
\includegraphics[width=2.0in]{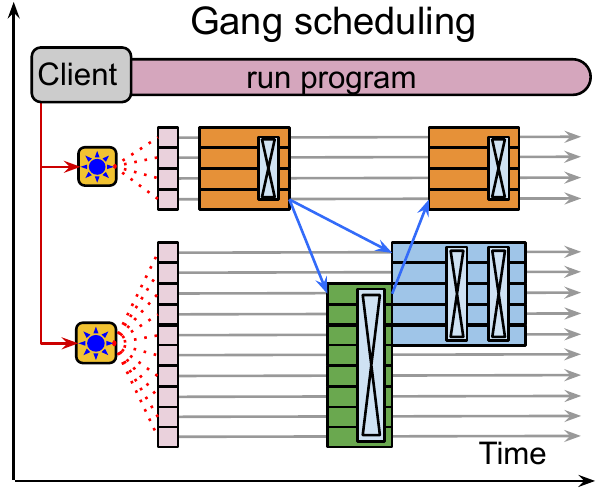}
\captionshrink
\caption{\ourname system overview. %
 (Left) Distributed computation expressed as a DAG where each node represents an individual compiled function, and edges between nodes represent data flows between functions. (Middle) Resource Manager allocates subsets of an island's accelerators ("virtual slices") for each compiled function.  %
 (Right) Centralized schedulers for each island gang-schedule computations 
 that are then dispatched by per-shard executors. Red arrows indicate control messages, blue arrows show data-path transfers.
}
\label{fig:system_overview}
\vspace{-2ex}
\end{figure*}

\

\ourname builds extensively on prior systems, including XLA~\cite{XLA} to represent and execute TPU computations, TensorFlow graphs and executors~\cite{abadi2016tensorflow} to represent and execute distributed CPU computations, and Python programming frameworks including JAX~\cite{jax2018github} and TensorFlow APIs. 
By leveraging these building blocks we are able to focus on the novel coordination aspects of \ourname, while being able to run existing ML models with minimal code changes.

\subsection{Resource Manager}\label{sec:coordinator}
A \ourname backend consists of a set of accelerators grouped into tightly-coupled islands that are in turn connected to each other over DCN (Figure~\ref{fig:system_overview}). \ourname has a ``resource manager'' which is responsible for the centralized management of devices across all of the islands.
A client may ask for ``virtual slices'' of the island with specific 2D or 3D mesh shapes that suit their communication pattern.
Each virtual slice contains ``virtual devices`` that allow the client to express how computations are laid out on the mesh.
The resource manager dynamically assigns physical devices for virtual devices satisfying the desired interconnect topology, memory capacity, etc.

Our initial resource manager implementation uses a simple heuristic that attempts to statically balance load by spreading computations across all available devices, and keeps a one to one mapping between virtual and physical devices. 
If future workloads require it we can adopt a more sophisticated allocation algorithm, for example taking into account the resource requirements of all client computations and the current state of the system to approximate an optimal allocation of physical devices to computations.

\ourname allows backend compute resources to be added and removed dynamically, with the resource manager tracking available devices. The layer of indirection between virtual and physical devices, as enabled by our single-controller design, will allow us in future to support features like transparent suspend/resume and migration, where a client's virtual devices are temporarily reclaimed or reassigned without the need for cooperation from the user program.

\subsectionshrinker
\subsection{Client}\label{client}
When the user wants to run a traced program, it calls the \ourname client library which first assigns virtual devices to any computations that have not been run before, and registers the computations with the resource manager, triggering the servers to compile the computations in the background. The client then constructs a device location-agnostic \ourname intermediate representation (IR) for the program, expressed as a custom MLIR~\cite{mlir} dialect. The IR is progressively ``lowered'' via a series of standard compiler passes, which eventually output a low-level representation that includes the physical device locations. This low-level program takes into account the network connectivity between physical devices and includes operations to transfer outputs from a source computation shard to the locations of its destination shards, including scatter and gather operations when a data exchange is required. It is efficient to repeatedly run the low-level program in the common case that the virtual device locations do not change, and the program can be re-lowered if the resource manager changes the mapping between virtual and physical devices.

The client in older single controller systems can quickly become a performance bottleneck as it coordinates thousands of individual computations and data buffers corresponding to each shard of computations spread across thousands of accelerators. The \ourname client uses a \emph{sharded buffer} abstraction to represent a logical buffer that may be distributed over multiple devices. This abstraction helps the client scale by amortizing the cost of bookkeeping tasks (including reference counting) at the granularity of logical buffers instead of individual shards.

\subsectionshrinker
\subsection{Coordination implementation}

\ourname relies on \plaquename for all cross-host coordination that uses DCN. \plaquename  is an existing (closed-source) production sharded dataflow system used at \google for many customer-facing services where high-fanout or high-fanin communication is necessary, and both scalability and latency are important. The low-level \ourname IR is converted directly to a \plaquename program, represented as a dataflow graph. \ourname has stringent requirements for its coordination substrate, all of which are met by \plaquename.

First, the representation used to describe the \ourname IR must contain a single node for each sharded computation, to ensure a compact representation for computations that span many shards, i.e. a chained execution of 2 computations $A$ and $B$ with $N$ computation shards each should have 4 nodes in the dataflow representation: $Arg \xrightarrow{} Compute(A) \xrightarrow{} Compute(B) \xrightarrow{} Result$, regardless of the choice of $N$. 
In the \plaquename runtime implementation each node generates output data tuples tagged with a destination shard, so when performing data-parallel execution $N$ data tuples would flow, one between each adjacent pair of IR nodes.

The coordination runtime must also support \emph{sparse} data exchanges along sharded edges, in which messages can be sent between a dynamically chosen subset of shards, using standard progress tracking mechanisms~\cite{akidau2013millwheel,murray2013naiad} to detect when all messages for a shard have been received. Efficient sparse communication is a requirement to avoid the DCN becoming a bottleneck for data-dependent control flow on accelerators, which is one of the key capabilities that we want \ourname to enable.

\begin{figure*}[t]
    \centering
    \begin{subfigure}[t]{0.48\linewidth}
    \centering
    \includegraphics[scale=0.70]{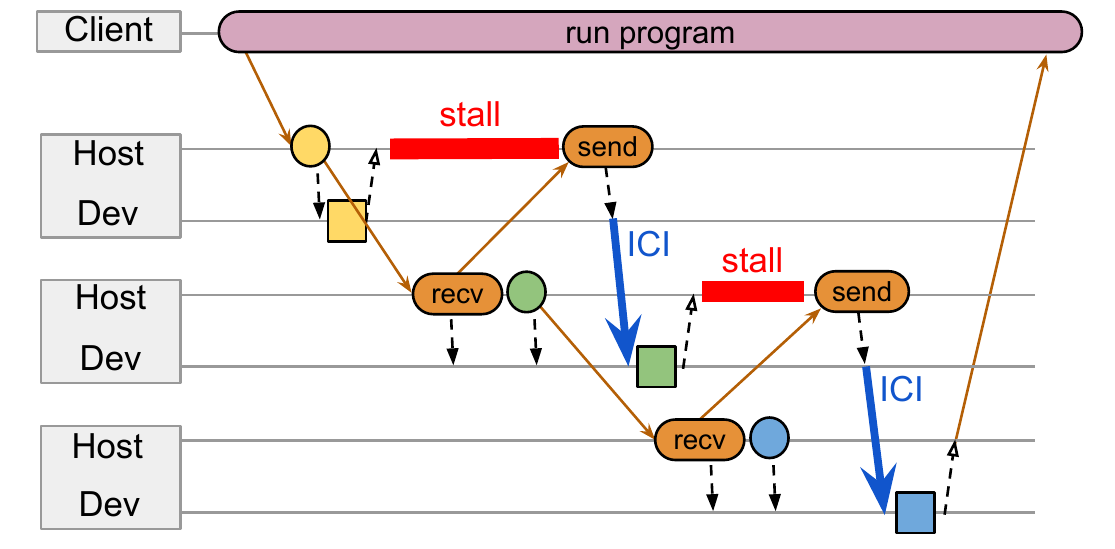}\\
    \subcaption{{\small Sequential dispatch}}
    \label{subfig:sequential-dispatch}
    \end{subfigure}
    \hspace{1mm}
    \begin{subfigure}[t]{0.48\linewidth}
    \centering
    \includegraphics[scale=0.70]{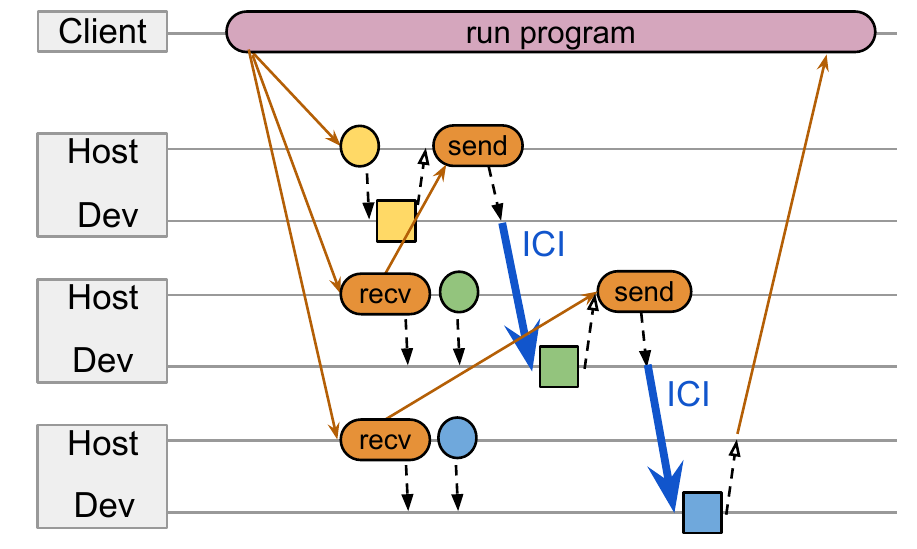}\\
    \subcaption{{\small Parallel dispatch}}
    \label{subfig:parallel-dispatch}
    \end{subfigure}
    \captionshrink
    \caption{Sequential vs.\ Parallel dispatch for a three-node program. When a computation's on-device execution time is \emph{shorter} than the time spent in scheduling, resource allocation, and coordination, the asynchronous pipeline stalls because of host-side work in sequential dispatch. Parallel asynchronous dispatch overcomes this bottleneck by running the host-side work in parallel, exploiting the statically known resource usage of regular compiled functions. Schedulers are omitted for brevity. %
    }
    \label{fig:parallel_dispatch}
    \vspace{-2ex}
\end{figure*}

The coordination substrate is used to send DCN messages that are in the critical path for transmitting scheduling messages and data handles (Figure~\ref{fig:parallel_dispatch}), so it must
send critical messages with low latency, and batch messages destined for the same host when high throughput is required.

It is also convenient to use an extensible, general-purpose, dataflow engine to handle DCN communication, since this means that \ourname can also use it for background housekeeping tasks such as distributing configuration information, monitoring programs, cleaning them up, delivering errors on failures, and so on.

We believe that it would be feasible to re-implement the full \ourname design using other distributed frameworks such as Ray~\cite{moritz2018ray} rather than \plaquename to realize the low-level coordination framework. In such an implementation, \ourname executors and schedulers would be replaced by long-running Ray actors that would implement \ourname scheduling on top of the underlying Ray cluster scheduling, and executors could use PyTorch for GPU computation and collectives. Some additions might be required to attain comparable performance (see \S\ref{sec:evaluation}) because Ray lacks, for example, an HBM object store, or primitives to efficiently transfer remote objects over the GPU interconnect.

\subsectionshrinker
\subsection{Gang-scheduled dynamic dispatch}\label{sec:impl:gangsched}

As discussed previously (\S\ref{sec:singlevsmulti}), one requirement for supporting SPMD computations on a shared set of accelerators is to support efficient gang-scheduling.
The \ourname runtime includes a centralized scheduler per island that consistently orders all of the computations in the island. %
As \ourname enqueues a program for execution, the \plaquename dataflow program is responsible for \emph{(i)} enqueueing the execution of local compiled functions at each accelerator, with buffer futures as inputs; \emph{(ii)} enqueueing network sends to remote accelerators for the buffer futures output by function executions;  and \emph{(iii)} communicating with the scheduler to determine a consistent order of function executions across all programs running on the island.  The scheduler must implement policies for allocating accelerators at a time-scale of milliseconds. Our current implementation simply enqueues work in FIFO order, but more sophisticated schedulers might for example reorder computations based on estimated execution times.

\sectionshrinker
\subsection{Parallel asynchronous dispatch}\label{sec:async}

When running computations on accelerators, systems can take advantage of asynchronous APIs to overlap computation with coordination~\cite{kwon2020nimble}.
Consider the three-node graph in  Figure~\ref{subfig:sequential-dispatch}, where the squares correspond to three nodes A, B, and C running on accelerators attached to hosts A, B, and C. All node computations are regular compiled functions. Host A enqueues node A, receives a future for A's outputs, and transmits the future to host B. Host B allocates B's inputs, transmits the input buffer addresses to host A, and performs most of the preparatory work to launch node B's function. When node A completes, its outputs are sent via the accelerator interconnect directly into node B's input buffers, and then host B starts node B. The latency between one node completing and the next node starting can be made to be little more than the data transfer time.

The above design works well when a predecessor node's computation takes \emph{longer} than the time spent in scheduling, resource allocation, and coordination between hosts. However if the computation time is too short, which is the case shown in the figure, the asynchronous pipeline stalls and the host-side work becomes the critical bottleneck for executing the overall sequence of computations. Given that the compiled functions are all regular, a successor node's input shapes can in practice be computed \emph{before} the predecessor computation was even enqueued. 

We therefore introduce a novel \emph{parallel asynchronous dispatch} design shown in Figure~\ref{subfig:parallel-dispatch}, which exploits the statically known resource usage of regular compiled functions to run most of the host-side work for a computation's nodes in parallel, rather than serializing the work for a node to happen after its predecessors have been enqueued.
Since work can only be scheduled in parallel when functions are regular, \ourname %
treats parallel scheduling as an optimization and falls back to the traditional model when a node's resource requirements are not known until a predecessor computation has completed (e.g.,\ due to data-dependent control flow). When a subgraph of a computation can be scheduled statically, the program sends a single message (describing the entire subgraph) to the scheduler, which is able to sequence the execution of all the active shards in the subgraph back to back. The use of a single message is designed to minimize network traffic, but does not require the scheduler to actually \emph{enqueue} all the subgraph's shards as a batch: computations may still be interleaved with those 
submitted by other concurrently executing programs.
We evaluate the cost of different dispatch mechanisms in~\S\ref{sec:evaluation}.

\subsectionshrinker
\subsection{Data management}

Each host manages a \emph{sharded object store} that is similar to Ray's object stores~\cite{moritz2018ray}, but extended to also track buffers held in accelerator HBM at each shard. Client programs can hold references to objects in remote host or accelerator memory, and the client and servers refer to them using opaque handles that allow the system to migrate them if needed. Intermediate program values are also kept in the object stores, for example while the system is waiting to transfer them between accelerators, or pass them to a subsequent computation. The objects are tagged with ownership labels so that they can be garbage collected if a program or client fails. We can use simple back-pressure to stall a computation if it cannot allocate memory because other computations' buffers are temporarily occupying HBM.

\begin{figure}[t]
    \centering
    \includegraphics[width=.9\linewidth,clip,trim=2mm 0mm 2mm 2mm]{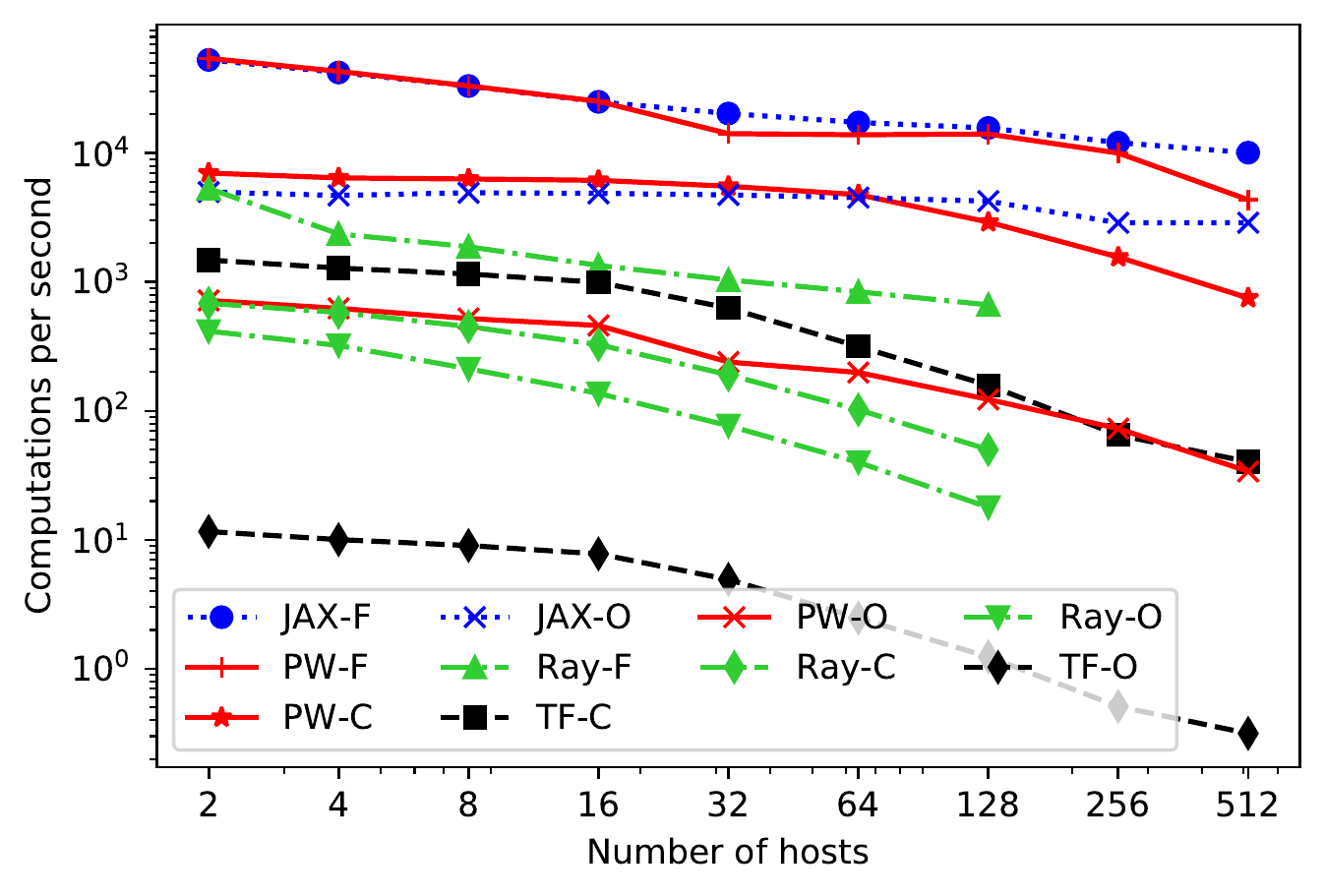}
    \captionshrink
    \setlength{\abovecaptionskip}{-3pt}
    \caption{Dispatch overhead of \ourname compared to TF, JAX, and Ray. \ourname outperforms single-controller systems like TF and Ray on all configurations, and matches the performance of multi-controller JAX in Fused~(-F) and Chained~(-C) configurations for up to 1000 and 256 TPU cores, respectively. Each computation comprises a single scalar AllReduce followed by a scalar addition.}
    \label{fig:bench0plus1}
    \vspace{-2ex}
\end{figure}

\sectionshrinker
\section{Evaluation}\label{sec:evaluation}

For evaluating JAX, \ourname, and TensorFlow on TPU we use three different configurations. Configuration~(\textbf{A}) has $4$ TPUs per host, and the largest instance we report on has $512$ hosts, resulting in $2048$ total TPUs connected via ICI\@. Configuration~(\textbf{B}) has $8$ TPUs per host, and the largest instance we report on has $64$ hosts, and a total $512$ TPUs. Configuration~(\textbf{C}) uses four islands of TPUs, where each island has $4$ hosts and $32$ TPUs. We note in the text when experiments use a subset of the TPUs of a particular configuration. 

When evaluating Ray on GPU we use Ray v1.3 and PyTorch 1.8.1 running on \texttt{p3.2xlarge} VMs\footnote{These VMs have $1\times$V100 GPU and $8\times$CPU cores.} with hosts connected via DCN and scheduled using Amazon placement groups.

We mostly compare \ourname against multi-controller JAX, since JAX has demonstrated state of the art performance in industry standard benchmarks~\cite{mattson2020mlperf} and we can easily run JAX and \ourname~(PW) on identical hardware configurations. We also compare against TensorFlow~(TF) and Ray in micro-benchmarks, to examine specific aspects of \ourname's distributed system performance,
and show pipelined performance of a TF model running on \ourname.

\begin{figure}[t]
    \centering
    \includegraphics[width=.9\linewidth,clip,trim=5mm 0mm 15mm 12mm]{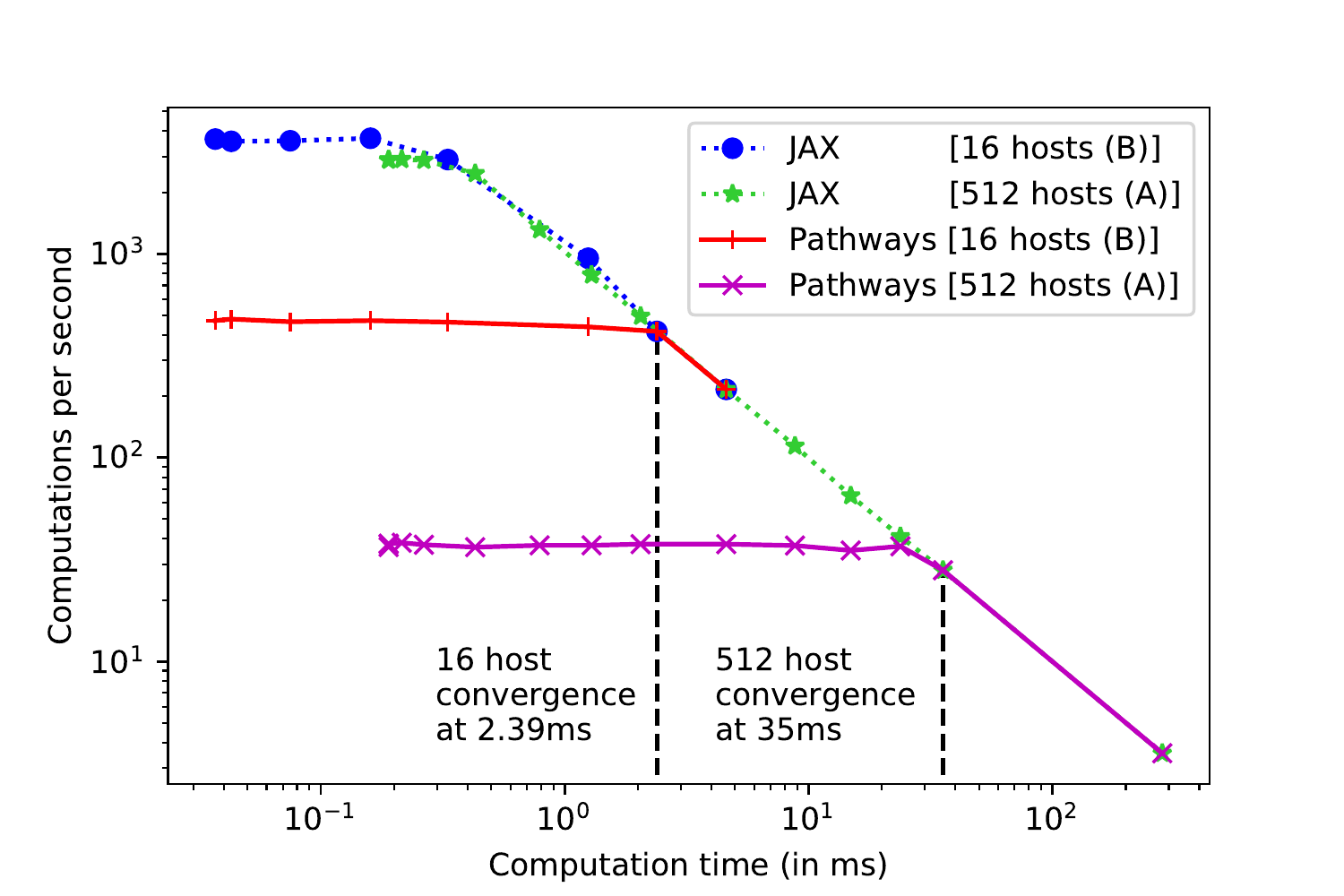}
    \captionshrink
    \setlength{\abovecaptionskip}{-2pt}
    \caption{Smallest computation to match throughput between \ourname and JAX, masking the single-controller overhead. \ourname matches JAX throughput for a computation size of at least 2.3~ms for 16 hosts with 128 TPUs on configuration~(B), and for a computation size of at least 35~ms for 512 hosts with 2048 TPUs on configuration~(A).}
    \label{fig:bench0matchingthroughput}
    \vspace{-2ex}
\end{figure}

\subsectionshrinker
\subsection{Single-controller dispatch overheads}

Our first experiment is a micro-benchmark to compare the overheads of JAX multi-controller with single-controller frameworks. We construct programs that repeatedly run a trivial gang-scheduled computation containing a single AllReduce of a scalar followed by a scalar addition, feeding the output of one computation to the input of the next. We measure the throughput: the number of computations per second that execute on the accelerators. We compare three ways that the user code can enqueue the 
computations:
\begin{itemize}[noitemsep,topsep=0pt,parsep=0pt,partopsep=0pt,leftmargin=*]
    \item \textbf{OpByOp~(-O):} The user code contains a separate call for each execution of the computation.
    \item \textbf{Chained~(-C):} The user code contains a series of calls each of which executes a chain of $128$ nodes, where each node executes the computation. The system executes the entire chain
    in response to a single client call.
    \item \textbf{Fused~(-F):} The user code contains a series of calls each of which executes a single computation node, where the node contains a chain of $128$ computations.
\end{itemize}
For JAX multi-controller, OpByOp means JIT-compiling a function containing one computation and calling it repeatedly from Python, and Fused means JIT-compiling a function containing a chain of computations. There is no analog of Chained for a multi-controller. For \ourname, OpByOp and Fused use the same JAX source as for the multi-controller, and Chained uses the \ourname program tracer to form a multi-node program where each node contains a simple computation. TF is similar to \ourname, where we construct the same TPU computations and execute them using TF graphs instead of \ourname. For Ray, OpByOp means executing a separate actor method for each computation which executes a PyTorch AllReduce. Chained means chaining a sequence of actor methods (by passing Ray futures), each of which executes a single PyTorch AllReduce. Fused means executing a single actor method which runs a chain of PyTorch AllReduce commands in a loop.

\begin{figure}[t]
\vspace{-1.8ex}
\centering
\begin{minipage}[t]{\linewidth}
    \begin{figure}[H]
    \centering
    \includegraphics[width=.8\linewidth,clip,trim=3mm 1mm 15mm 12mm]{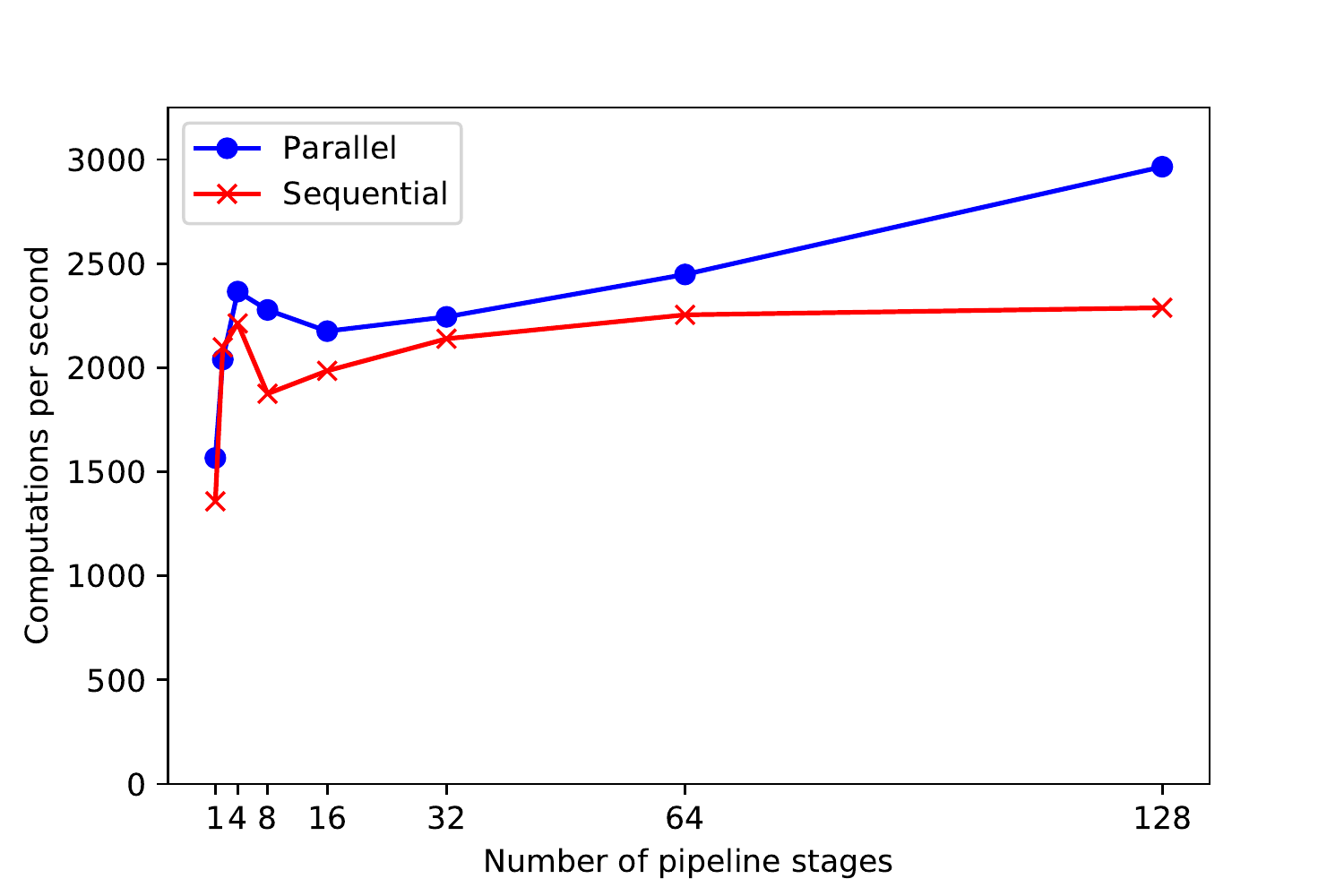}
    \captionshrink
    \setlength{\abovecaptionskip}{-2pt}
    \caption{Parallel vs. Sequential Async Dispatch in \ourname. Each pipeline stage runs on a different set of 4 TPU cores (different host) transferring data to next stage via ICI. With parallel async dispatch, \ourname amortizes the fixed client overhead and the scheduling overhead for  large number of pipeline stages.} %
    \label{fig:par_seq}
    \end{figure}
\end{minipage}
\vspace{-2ex}
\end{figure}

Figure~ shows the result. Note that OpByOp is a worst-case experiment that is not idiomatic for any of the frameworks, and it is present merely to stress the underlying systems. As expected, for OpByOp the JAX multi-controller throughput is much better than the single-controller systems, particularly as the number of accelerators increases. Most of \ourname's overhead comes from the fact that the client waits until the coordinator has enqueued one computation and returned its output handles before enqueueing the next. We could eliminate most of this overhead by allowing user code to proceed in parallel with the enqueue RPC, and opportunistically batching multiple small computations into a single \ourname program. We have not focused on optimizing overheads of very small computations since, on real models with computations involving more than scalars, \ourname already matches the performance of multi-controller JAX (see \S\ref{sec:large_scale_models}). Once enough work is Fused into a single node \ourname matches JAX's performance up to 1000 TPU cores, and \ourname Chained outperforms JAX OpByOp up to 256 cores, because \ourname can execute back-to-back accelerator computations directly from C++ while JAX OpByOp transitions to Python for every computation.

TensorFlow and Ray suffer from their lack of a device object store: Ray must transfer the result of a computation from GPU to DRAM before returning the object handle to the client, while TensorFlow transfers the \emph{data} back to the client. This overhead hurts their OpByOp performance but is largely amortized for Chained and Fused. The performance of Ray and \ourname are not directly comparable since they use different hardware, but we interpret the results to suggest that, if the full \ourname design were implemented substituting Ray for \plaquename, it should be possible to achieve comparable performance. Out of the box, Ray shows about an order of magnitude worse performance per computation than \ourname, but that is unsurprising since Ray can execute general-purpose Python actors and \ourname is specialized to TPU computations launched from C++. With careful attention to engineering, it might be possible to add fast paths to Ray, such as an on-GPU object store and primitives to transfer objects efficiently over the GPU interconnect, that eliminate most of its additional overheads. TensorFlow is slow when running over many cores because it uses a centralized barrier, implemented with control edges, to serialize the gang-scheduled computations. 

Figure~\ref{fig:bench0matchingthroughput} varies the amount of time spent in each computation to find the smallest computation for which \ourname matches JAX's throughput. For 16 hosts with 128 TPUs on configuration~(B), parity is reached with only $2.3$\,ms, and even for 512 hosts with 2048 TPUs on configuration~(A), a computation of at least $35$\,ms masks all of \ourname's single-controller overhead.

Our next micro-benchmark, also on configuration~(B), evaluates the benefit of the parallel asynchronous dispatch mechanism described in \S\ref{sec:async}. We construct a more realistic pipeline benchmark in which the simple computations from the earlier benchmark are again chained together, but now each computation runs on a different set of 4 TPU cores, each on a different host, and data output from one computation must be sent via ICI before the next computation can execute.
Figure~\ref{fig:par_seq} shows 
three ``phases'': at first the fixed client overhead is amortized as the number of hosts increases; then the increasing transfer costs of adding more stages begin to dominate; finally the system starts to amortize the fixed scheduling overhead. Eventually we expect that transfer overheads would dominate again. For comparison, we also show the performance when we force the \ourname dataflow execution to use sequential asynchronous dispatch, and wait for one computation to be enqueued before enqueueing the next, to measure the benefit we get from parallel asynchronous dispatch.

\begin{figure}[t]
\centering
    \centering
    \includegraphics[width=.8\linewidth,clip,trim=5mm 2mm 15mm 11mm]{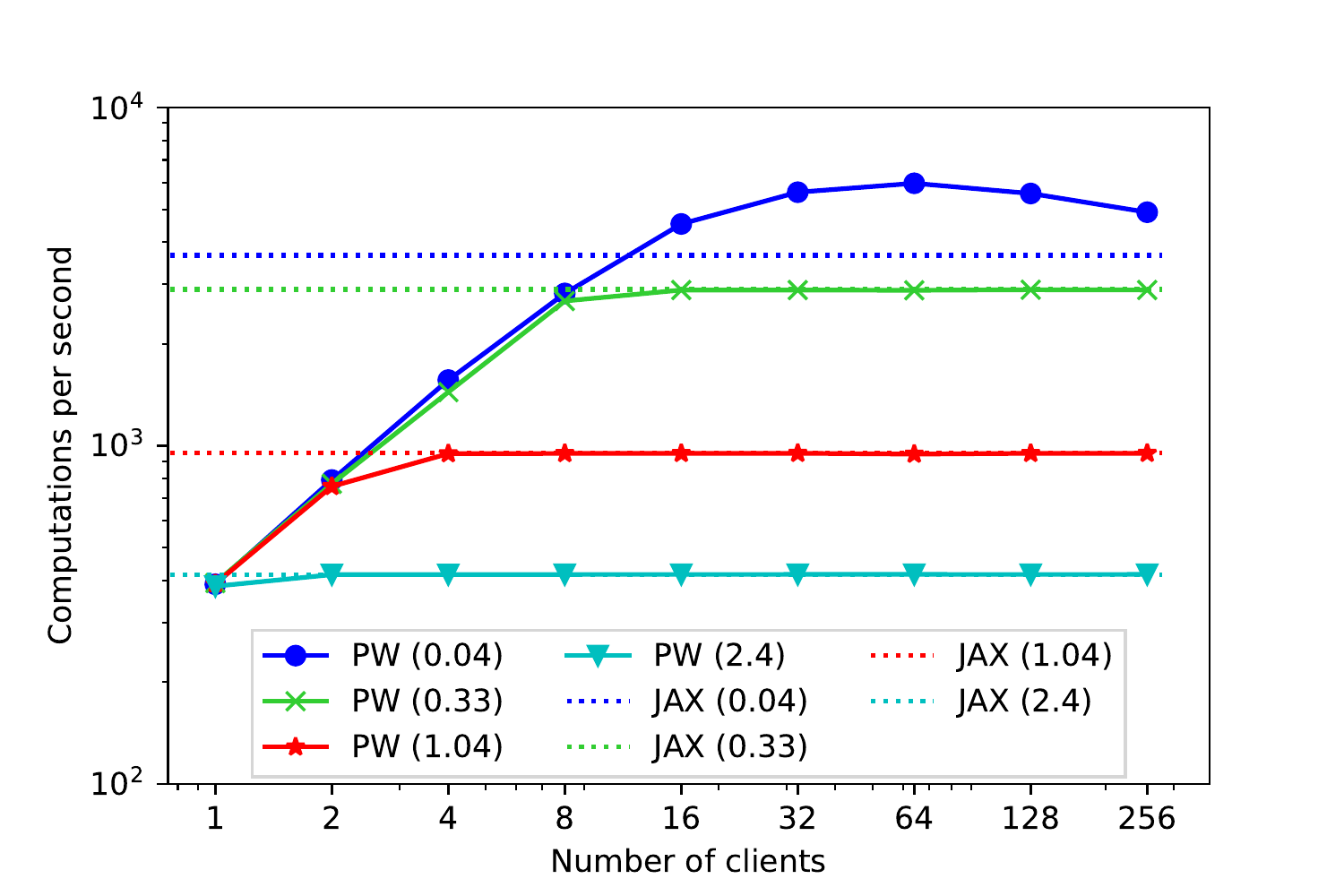}
    \captionshrink
    \caption{Aggregate throughput of concurrent programs (compute times in ms). \ourname time-multiplexes accelerators between programs efficiently incurring no overhead to context switch.
    }
    \label{fig:bench0concurrent}
\vspace{-2ex}
\end{figure}

\subsectionshrinker
\subsection{Multi-tenancy}
\label{sec:eval-multi-tenancy}

We validate in Figure~\ref{fig:bench0concurrent} (performed on configuration~(B)) that \ourname is able to time-multiplex accelerators between concurrent programs. \ourname can achieve at least the same aggregated throughput as JAX when multiple clients concurrently submit \emph{different} \ourname programs, i.e., there is no overhead to context switch between programs from different clients, at least when their resources concurrently fit in HBM (traces in Appendix~\ref{sec:app:additional_result}). As already shown in Figure~\ref{fig:bench0matchingthroughput}, the degree of concurrency required to match the throughput is lower for larger computation sizes since the TPU cores reach full utilization sooner.
It is noteworthy that
the maximum throughput of \ourname exceeds that that of JAX for very small computations,
achieving higher TPU utilization.
This is because a \ourname worker can accept more computations from remote clients than JAX can dispatch using Python locally.

\begin{figure}
    \centering
    \begin{subfigure}{0.75\linewidth}
        \centering
        \includegraphics[width=.85\linewidth,clip,trim=0px 25px 0px 0px]{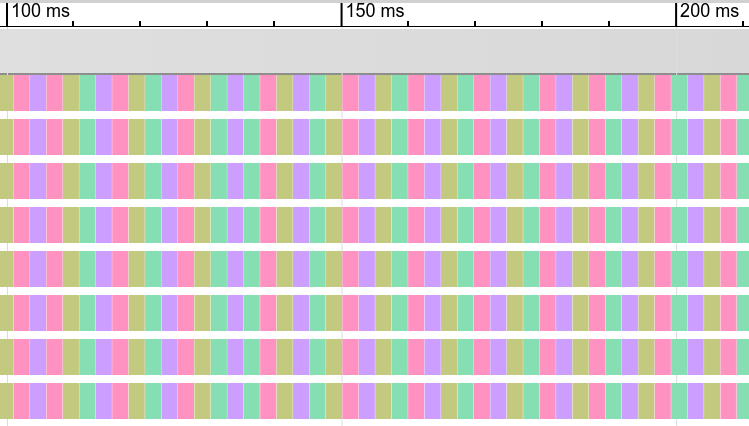}\\
    \end{subfigure}
    \begin{subfigure}{0.75\linewidth}
        \centering
        \includegraphics[width=.85\linewidth,clip,trim=0px 24px 0px 0px]{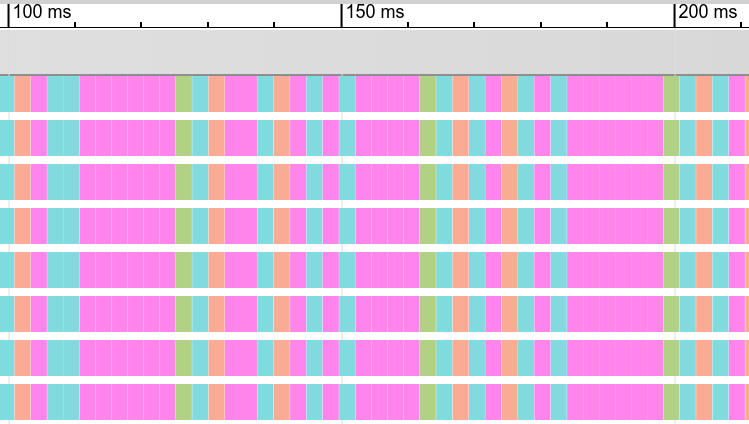}\\
    \end{subfigure}
    \captionshrink
    \caption{Traces of a sample of cores on \ourname showing interleaving of gang-scheduled concurrent programs with proportional-share ratios of 1:1:1:1 (Upper) and 1:2:4:8 (Lower) between 4 clients.}
    \label{fig:bench0concurrent-trace}
\vspace{-2ex}
\end{figure}

Figure~\ref{fig:bench0concurrent-trace} shows traces of a sample of 128 cores on \ourname for the above workload.
This experiment highlights that \ourname performs gang-scheduling of programs submitted by 4 independent clients while controlling allocation of accelerator time for fairness;
for example, the scheduler can enforce proportional share in this multi-tenancy setting.

\subsectionshrinker

\subsection{Large scale model performance}
\label{sec:large_scale_models}
Finally, we show the performance of \ourname in training real machine learning models that can be expressed as SPMD programs. We compared JAX and TF models running on their native systems to the same models running on \ourname, and verified that at numerical results are identical, so we focus only on performance.

We first compare to JAX multi-controller running a Transformer model with an Encoder-Decoder architecture that is used for several text-to-text natural language processing tasks. We use model configurations from~\cite{raffel2019exploring} and run the experiments on TPUv3s with 16GB memory per accelerator. Table~\ref{table:tput_t5_model} shows the training throughput (tokens/second) for Text-to-text Transformer model with various model sizes (up to 11 billion parameters), training on different number of accelerators. As expected, since the model code is the same, the models trained on JAX and \ourname achieve the same perplexity in the same number of steps. Over all tested model sizes, the two systems show identical performance since realistic computations are large enough to mask single-controller overheads. While we do not report detailed results, we have substantial experience of running JAX models on \ourname, which corroborates the finding that the performance of the two systems is comparable across a broad range of settings.

\begin{table}[t]
\small
\begin{center}
\caption{Training throughput (tokens/s) of Text-to-text Transformer model configurations from~\cite{raffel2019exploring} on JAX multi-controller and \ourname.}
\label{table:tput_t5_model}
\vspace{0.08in}
\begin{tabular}{l r r r r }
\hline
Model & Params & TPU cores & JAX & \ourname \\
\hline
T5-Base & 270M & 32 & 618k &  618k\\
T5-Large & 770M & 32 & 90.4k & 90.4k  \\
T5-3B & 3B & 512 & 282.8k & 282.8k \\
T5-11B & 11B & 512 & 84.8k & 84.8k \\
\hline
\end{tabular}
\end{center}
\vspace{-2ex}
\end{table}

Next, we compare the performance of \ourname when training a Transformer-based language model with a Decoder-only architecture on configurations~(B) and~(C). For this experiment, we use a model expressed in Python using TF. The model consists of 62 Transformer layers with a model dimension of 2048 and a hidden dimension of 8192, which results in 3 billion parameters in total. We compare an SPMD configuration to a pipeline using a GPipe-like schedule~\cite{huang2019gpipe}. The pipelined model is split into multiple stages with balanced computation in each stage. Since the first stage has an extra embedding lookup layer and the last stage has an extra softmax layer, we took out one Transformer layer from the first and last stage to balance the amount of compute per stage. Each stage is assigned to a different set of accelerators spanning multiple hosts.

Table~\ref{table:pipeline_pw_model} shows the training throughput for different numbers of stages (S) and micro-batches (M), while keeping the global batch size and training hyperparameters fixed.\footnote{Unlike Megatron~\cite{shoeybi2019megatron}, the SPMD-sharded model evaluated here is similar to GShard~\cite{lepikhin2020gshard} and does \emph{not} have communication proportional to batch size, so it is fair to evaluate pipelined and SPMD with the same batch size.} The number of examples per micro-batch is fixed at 4 for all cases, and, hence, the global batch size per step is 2048 for the 128-core configurations (8192 for the 512-core one). 

\ourname's training throughput increases %
proportionally with the number of TPU cores per pipeline stage (Table~\ref{table:pipeline_pw_model}),
in line with other systems~\cite{rasley2020deepspeed,narayanan2021efficient}.
This result is consistent with Figure~\ref{fig:bench0plus1} showing that the throughput of \ourname linearly scales with the number of hosts.
Increasing the number of pipeline stages adds minimal overhead, the throughput being reduced from 133.7k tokens/sec to 131.4k tokens/sec when the number of stages increases from 4 to 16. We compare the pipelined models' performance to an equivalent model expressed using SPMD, and observe that at least in this instance, the pipeline has competitive performance to SPMD, since collective communication within the SPMD computation incurs higher overhead than pipeline bubble overhead.

\begin{figure}[t]
\vspace{-1.2ex}
\centering
\begin{minipage}[t]{\linewidth}
    \begin{figure}[H]
    \centering
    \includegraphics[width=\linewidth]{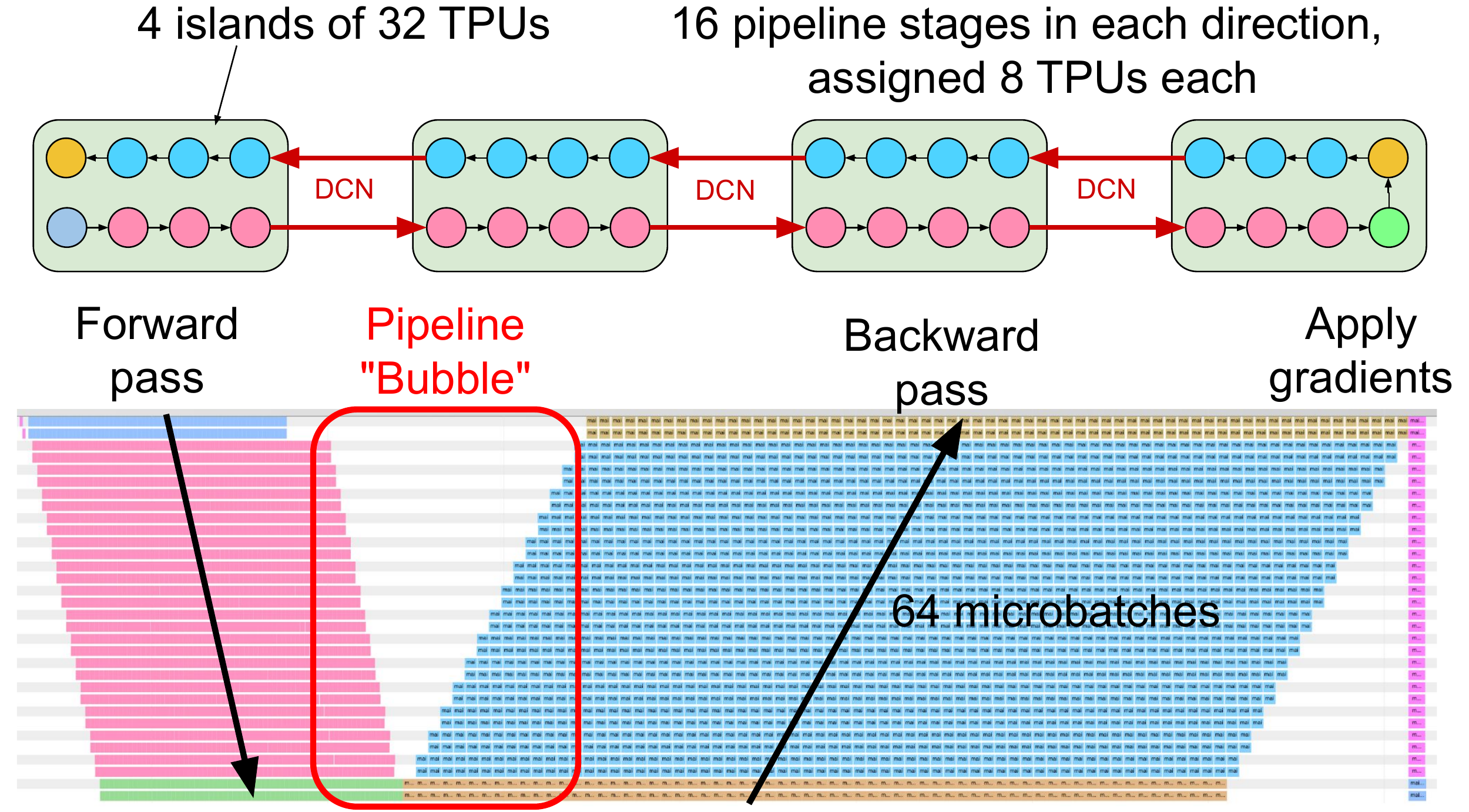}
    \captionshrink
    \caption{3B Transformer model pipelined over 128 TPUs: \ourname can  efficiently  train models over islands of TPUs connected via DCN achieving the same throughput (131.4k tokens/sec) on 4 islands of 32 cores each on configuration~(C) as using a single island of 128 cores on configuration~(B).}
    \label{fig:gpipe-xprof}
    \end{figure}
\end{minipage}
\vspace{-2ex}
\end{figure}

We also demonstrate that \ourname can efficiently train models over islands of TPUs connected via DCN. In the $S=16,\ M=64$ configuration with 128 cores, we measure the same throughput ($131.4$k tokens/sec) using a single island of 128 cores on configuration~(B), or 4 islands of 32 cores each on configuration~(C).
Figure~\ref{fig:gpipe-xprof} shows a trace of a sample of cores when the stages are partitioned into islands. DCN transfers occur between every group of 8 rows in the trace, and are not visible in the trace because communication time is effectively overlapped with computation.

\begin{table}[t]
\vspace{-1ex}
\small
\begin{center}
\caption{Training throughput (tokens/s) of 3B Transformer language model, using SPMD or multiple pipeline stages, with $C$ TPU cores in \ourname. For pipeline-parallel models, there are $S$ stages and each batch is split into $M$ $\mu$-batches.}
\label{table:pipeline_pw_model}
\vspace{0.05in}
\begin{tabular}{l r r}
\hline
Model configuration & TPU cores & \ourname \\
\hline
Model-parallel (SPMD) & 128 & 125.7k \\
Pipelining, S=4, M=16 & 128 & 133.7k \\
Pipelining, S=8, M=32 & 128 & 132.7k \\
Pipelining, S=16, M=64 & 128 & 131.4k \\
Pipelining, S=16, M=64 & 512 & 507.8k \\
\hline
\end{tabular}
\end{center}
\vspace{-2ex}
\end{table}

Finally, we scale up training of large Decoder-only Transformer models to 64B and 136B parameters using two islands of accelerators.
When trained using using \ourname over \emph{two} islands of compute connected over DCN, \ourname achieves $\sim97\%$ of the throughput as compared to a single island with twice as many devices. For the 136B (64B) LM model, we train over two islands of 1024 (512) cores that uses the fast ICI within island reduction followed by DCN transfer across islands (execution trace available in Appendix~\ref{sec:app:additional_result})  of 1030GB (457GB) for global reduction. 

\section{Discussion}

\subsectionshrinker
\subsection{\ourname{} design vs. implementation}\label{sec:designvsimpl}

\ourname was designed to target large collections of TPU accelerators. The use of TPU instead of GPU affects many of our low-level design decisions. The biggest difference between TPU and GPU is that far longer-running and more complex computations can be fused into a single TPU kernel, because the TPU supports rich control flow and communication primitives that must instead be executed by driver code on GPU systems. GPUs, by contrast, are more tightly integrated with host memory systems and DCNs~\cite{gpudirect2021}
(more details in Appendix~\ref{sec:gpus-and-tpus}).
TPUs are a good fit for \ourname because XLA can compile high performance functions containing fused collectives, and the large islands of high-performance TPU interconnects allow flexible scheduling of computations of many different sizes. Nevertheless, we believe that most of the high-level architectural choices we made in \ourname and describe in this paper would also be valid for large-scale GPU systems.

\subsectionshrinker
\subsection{Resource management}\label{sec:resource-management}

\ourname is designed to allow a wide variety of fine-grained dynamic resource-management policies.
Our initial research has focused on efficient dynamic time-multiplexing of TPU computations\@.
For more complex future multi-tenancy use cases, \ourname will need to handle more diverse resource types including but not limited to device and host memory, and ICI, DCN, and PCIe bandwidth.
\ourname's single-controller model grants the system an extensive ability to track available resources and to allocate resources at large scale.
We are planning to explore common multi-tenancy requirements such as priorities, performance isolation, access control, and resource accounting, but 
at timescales that are significantly smaller than prior work, and for orders-of-magnitude larger pools of resources (e.g., thousands of cores and TBs of accelerator memory).%

\subsectionshrinker
\subsection{Data-dependent vectorized control flow}\label{sec:discussion-controlflow}

Almost all ML models currently update \emph{every} model weight based on every training example in every step. We want to enable research that uses fine-grain control flow so that different model weights can be updated per example, or even per sub-example (patch of an image, or word of a sentence). Models like Mixture of Experts (MoE)~\cite{shazeer2017outrageously} and routed capsule networks~\cite{hinton2018matrix,barham2019hotos} exploit computational sparsity by ``routing'' different (sub\nobreakdash-)examples to the accelerators hosting different subsets of model weights based on learned functions that are updated as training progresses. 
This routing requires fine-grain data-dependent data exchanges between nodes. Our ML research colleagues have told us that they would like to use sparsity more effectively when training ever larger models, with ever more tasks, but that current frameworks limit their ability to experiment with novel model architectures.
It is the subject of future work to support data-dependent vectorized control flow with both a clean programming model and good performance.

\sectionshrinker
\section{Related work}\label{sec:related}

We have examined closely related work in detail in~\S\ref{sec:singlevsmulti}. This section expands on related research that addresses ML workloads that need capabilities beyond those offered by SPMD multi-controllers, and validates our \ourname design choices.

Sharing accelerators across multiple tasks is crucial for achieving high resource utilization.
Conventional resource sharing is performed in a coarse-grained manner.
For example, general-purpose virtualization enables cloud applications to efficiently share multi-tenant resources with performance isolation~\cite{angel2014end, wentzlaff2010operating, shahrad2016availability, baumann2009multikernel}, but cloud providers dedicate accelerators to individual users. 
Cluster schedulers optimize for heterogeneity of ML workloads~\cite{narayanan2020heterogeneity} and multi-job, multi-user fairness and performance~\cite{xiao2018gandiva, ren2015hopper, mahajan2020themis, jeon2018multi}, but resources are still exclusively dedicated to single jobs at long time scales (seconds or more).

Recent work shows that finer-grained sharing can improve resource efficiency further:
virtualizing accelerators~\cite{yu2020ava, gupta2011pegasus, vijaykumar2016zorua} avoids dedicating a whole accelerator to a single user.
Large models~\cite{brown2020language} can be limited by available accelerator memory, requiring GPU memory virtualization~\cite{rhu2016vdnn, ausavarungnirun2018mask} or DRAM offload~\cite{rajbhandari2021zeroinfinity}.
Concurrent (time-multiplexed or overlapping) ML task execution~\cite{gupta2018shampoo,xiao2020antman,bai2020pipeswitch,yu2020fine,wang2021ticktock,lim2021zico} helps harvest idle resources within accelerators.
These fine-grained sharing techniques demonstrate opportunities for sharing accelerators that are hard to capitalize on at scale without a single-controller system like \ourname.

Many works have shown that deviating from SPMD computations can improve efficiency on large workloads:
pipelining~\cite{huang2019gpipe,narayanan2019pipedream,yang2021pipemare} partitions ML models into static heterogeneous computations across accelerators. Graph neural network training~\cite{jia2020improving}, neural architecture search~\cite{pham2018efficient}, and multi-modal multi-task learning systems~\cite{ma2018modeling, lepikhin2020gshard, zhao2019recommending} are examples of inherently heterogeneous and dynamic tasks that do not fit naturally in the SPMD model. We anticipate that upcoming large-scale efficient ML models may form a collection of shared layers and exclusive layers~\cite{bommasani2021foundation}, which are natural to express as MPMD\@.

\sectionshrinker
\section{Conclusions}\label{sec:conclusions}

\ourname matches state of the art multi-controller performance on current %
ML models which are single-tenant SPMD.
We have ensured strict compatibility with multi-controller JAX, and as we demonstrate in \S\ref{sec:evaluation}, \ourname matches JAX's performance across very large system scales, for all but the smallest computations.

At the same time, \ourname upends the execution model of JAX programs, pulling user code back into a single-controller model, and interposing a centralized resource management and scheduling framework between client and accelerators. The single-controller programming model allows users simple access to much richer computation patterns.
The resource management and scheduling layer permits the reintroduction of cluster management policies including multi-tenant sharing, virtualization and elasticity, all tailored to the requirements of ML workloads and accelerators. Our micro-benchmarks show interleaving of concurrent client workloads, and efficient pipelined execution, convincingly demonstrating that the system mechanisms we have built are fast and flexible, and form a solid basis
for research into novel policies to make use of them.

We have shown that careful system design and engineering lets us ``get the best of both worlds'', matching performance on today's ML models while delivering the features needed to write the models of tomorrow.

\section*{Acknowledgements}

We gratefully acknowledge contributions to the design and implementation of the \ourname system from many colleagues at Google, and from members of the wider machine learning community. We also thank Mart\'in Abadi, James Laudon, Martin Maas, and the anonymous MLSys reviewers for their helpful suggestions on the presentation of the work.
\\

\bibliographystyle{plainnat}
\bibliography{references}

\vfill
\clearpage

\appendix
\sectionshrinker
\section{Accelerator design considerations}\label{sec:background}

Hardware acceleration is critical to modern deep learning; unfortunately, achieving high performance with accelerators is a non-trivial systems exercise.
The following subsections list established techniques commonly employed in deep learning systems to achieve good performance.

\subsectionshrinker
\subsection{Batching}

Given the end of Dennard-scaling, accelerators implement hardware parallelism, often using SIMT~\cite{kirk2007cuda} or systolic array~\cite{jouppi2020tpu} designs.
While these hardware architectures remove the arithmetic bottleneck, memory bandwidth quickly becomes the critical resource, necessitating high-bandwidth memory (HBM), an expensive and limited-capacity memory technology.
Training schemes for modern neural networks leverage batching to unlock parallelism (good for feeding parallel ALUs) and enable memory re-use (a \texttt{float} is read from memory once and used for multiple computations, substantially reducing a computation's memory bandwidth needs).
Nevertheless, batching is not a panacea: it puts pressure on the limited HBM memory capacity, and very large batch sizes can slow model convergence rates~\cite{shallue2018measuring,you2017lars,lanchantin2020lamp,anil2021scalable}.
While modern GPUs support unified memory---a capability to transparently page memory between accelerators, or from HBM to the host's DRAM---if the user is not careful, an HBM-bandwidth bound computation could slow to PCIe bandwidth, dropping accelerator utilization by an order of magnitude~\cite{lim2021zico}.  %

\subsectionshrinker
\subsection{Asynchronous programming}

Accelerator abstractions rely on an asynchronous programming model to achieve performance; a synchronous abstraction wastes too many accelerator computation resources between PCIe latency, kernel scheduling overheads, and interrupt delays.
Computations are enqueued on \emph{streams} to be executed on the accelerator at some point in the future.
This asynchronous abstraction effectively masks dispatch latency for small operations, so long as a sufficiently large pipeline of work is maintained.

\subsectionshrinker
\subsection{High performance interconnects}

Modern deep neural networks are orders of magnitude larger than the capacity of accelerator (HBM) memory~\cite{lepikhin2020gshard, huang2019gpipe}.
The parallelism within these neural networks is amenable to sharding across multiple accelerators simultaneously, however high speed interconnects between accelerators then become critical for performance.
GPUs use interconnects such as NVLink for high-speed communication between ``islands'' of accelerators on a small number of hosts~\cite{naumov2020deep}, and use RDMA capabilities of ethernet and Infiniband NICs (GPUDirect) to rapidly communicate between the islands.
TPUs have a custom mesh network built directly into the chips, and chips can communicate directly without involving the host or the data-center network.
Dedicated GPU and TPU interconnects are typically exposed to applications via 30 year old MPI primitives (e.g., AllReduce) that must be gang-scheduled so that every program enters the same primitive at the same time.
As larger computations are run (e.g., training a larger neural network, or training a fixed-size neural network over more accelerators through a form of weak scaling called data-parallel scaling), faster collective operations and thus network bandwidth are required to maintain efficient utilization of aggregate cluster resources.
This has prompted significant experimentation with alternate chip-network topologies including hypercubes, and 2-D and 3-D mesh tori~\cite{naumov2020deep}.

\subsectionshrinker
\subsection{Single-tenancy}

Unlike most resources in a computer, accelerators are not often shared by multiple programs simultaneously.
Deep learning models can be easily scaled to use more memory by increasing parameter counts or batch sizes, and thus programs in practice consume most available accelerator (HBM) memory.
PCIe bandwidth is much smaller than HBM- or accelerator interconnect-bandwidth.
This means that fine-grained context-switching (where much of the data in HBM is paged out to host DRAM over PCIe) results in wasting a significant fraction of accelerator cycles.
Thus, when a host program is not fully utilizing an accelerator, the computational resources are stranded and cannot be used productively.
Further, preemption of accelerator resources is minimized in practice, resulting in sub-optimal resource scheduling in large, shared clusters serving heterogeneous workloads; it is difficult to allocate large quantities of physically proximate devices to take advantage of network locality.

\subsectionshrinker
\subsection{Contrasting GPUs and TPUs}
\label{sec:gpus-and-tpus}

While there are many similarities between GPUs and TPUs, there are some important differences.
GPU systems tend to have small islands of NVLink-connected devices (e.g., 8 GPUs within one host), with larger aggregations connected over infiniband or data-center networking technology. GPUs are typically programmed by dispatching many small pre-compiled ``kernels'' to the accelerator, and because they are pre-compiled, the kernels must support dynamic shapes. Any communication between GPUs, whether over NVLink or via DCN, is performed via the NCCL library and initiated by the host.

TPU systems have thousands of devices connected all-to-all, with hundreds of hosts per ``island''~(Figure~\ref{fig:system_overview} Middle).
TPUs contain a capable ``scalar core'' that coordinates the TPU's vector computation units, allowing a TPU to execute long-running functions written in XLA~\cite{XLA} without any host interaction, and these functions may include collective communication across the dedicated ICI network. Consequently, on TPU, an ML framework typically constructs a large XLA program, which is just-in-time (JIT) compiled and dispatched to the accelerator. The fact that a single XLA computation may run for orders of magnitude longer than a GPU kernel justifies increased optimization effort by the compiler such as static buffer assignment and automatic rematerialization of intermediate program values (saving memory capacity). As a consequence of this static buffer assignment, TPUs have only limited support for dynamic shapes, making them a good fit to the \ourname concept of regular compiled functions.

TPUs are restricted to run a single program at a time, with no local pre-emption, mostly because their high-performance RDMA communication implementation between devices makes safe pre-emption difficult without distributed coordination. Because computations are not pre-emptible, it is essential to enqueue communicating computations in a consistent order across devices, or the system will deadlock. This requirement translates to the necessity for \ourname to perform centralized gang-scheduling. As noted in the main text of the paper, however, gang-scheduling is also highly advantageous for GPU efficiency. For an cluster prioritizing ML training workloads, where throughput is more important than latency, it is more efficient to dedicate an entire GPU, or a static fraction of a GPU, to a single carefully sized computation at a time, than to allow the GPU driver and hardware runtime to dynamically multiplex its computational resources across competing concurrent computations. Therefore, even though GPUs can execute concurrent programs without centralized scheduling, there is still a benefit from using a design like \ourname to make more efficient use of resources.

\sectionshrinker
\section{Structure of a typical ML program}\label{sec:mlprogram}

This subsection describes a typical contemporary ML computation in terms
of the high level structure that maps sub-computations to accelerators,
and the lowering of a sub-computation to accelerator kernels.

The computations that are executed by an accelerator running an ML workload are dominated by what we call ``compiled functions''. These are sub-computations with the following characteristics:
\begin{itemize}[leftmargin=1.5em]
    \item Input and output types, and the shapes of any input/output tensors, are known before the input data have been computed.
    \item Bounds of any loops are either known when the node computation is scheduled, or specified as a maximum trip count with potential early termination.
    \item Conditionals are ``functional'' where both branches have the same output type, and resources are allocated in advance sufficient for either branch.
\end{itemize}
The constraints on compiled functions are mostly due to the co-evolution of ML models with hardware, discussed in detail in \S\ref{sec:background}. Here we discuss some of the implications of the fact that the resource requirements of compiled functions are known in advance.

Almost all of today's high performance ML computations are expressed as long stretches of compiled functions and only occasionally (if ever) branch based on data that is computed by a compiled function. Since the system can perform resource allocation for compiled functions in advance, contemporary ML frameworks exploit this property by enqueueing compiled functions asynchronously before their predecessors have run, allowing host-side work to be done in parallel with accelerator computations~\cite{jax2018github,paszke2019pytorch}.  Wherever possible the frameworks submit graphs of compiled functions to a ``just in time'' (JIT) compiler~\cite{chen2018tvm,XLA} that is able to exploit optimizations like layout assignment and fusion that can substantially improve the efficiency of the resulting accelerator code.

The need to optimize graphs of compiled functions to achieve peak accelerator performance
means that frameworks typically trace the execution of fragments of
high level (Python) code that can be lowered to compiled functions. Thus,
even though client code may be written in a high level language with
complex state bound to the running context at a host,
performance-sensitive node computations are often lowered to an
internal representation (IR) that is serializable and relatively easy
to send to a remote host for execution.

\begin{figure*}[th!]
\centering
\begin{subfigure}{0.40\linewidth}
    \centering
    \includegraphics[width=0.95\linewidth,clip,trim=0px 105px 0px 3px]{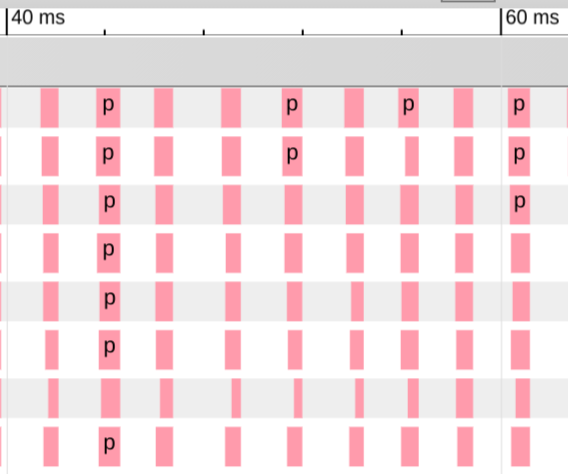}\\
    \subcaption{{\small 1 client}}
    \label{subfig:bench0concurrent-trace-1-short}
\end{subfigure}
\hspace{2em}
\begin{subfigure}{0.40\linewidth}
    \centering
    \includegraphics[width=0.95\linewidth,clip,trim=0px 105px 0px 2px]{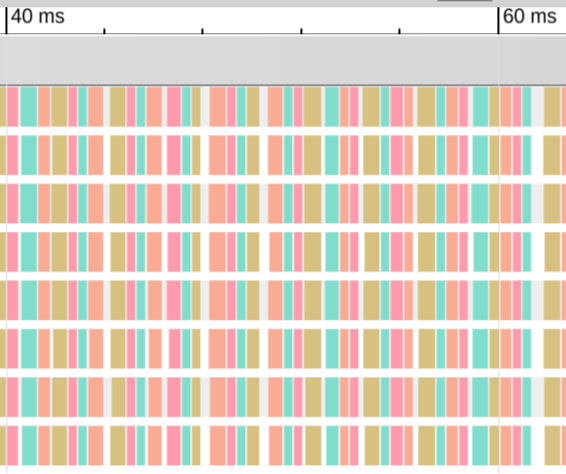}\\
    \subcaption{{\small 4 clients}}
    \label{subfig:bench0concurrent-trace-4-short}
\end{subfigure}
\vspace{1em}

\begin{subfigure}{0.40\linewidth}
    \centering
    \includegraphics[width=0.95\linewidth,clip,trim=0px 105px 0px 2px]{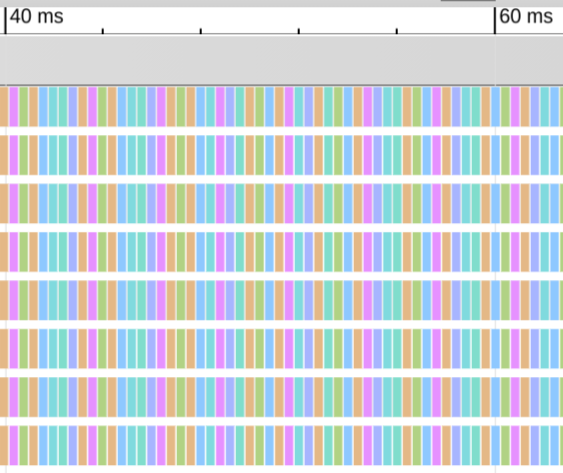}\\
    \subcaption{{\small 8 clients}}
    \label{subfig:bench0concurrent-trace-8-short}
\end{subfigure}
\hspace{2em}
\begin{subfigure}{0.40\linewidth}
    \centering
    \includegraphics[width=0.95\linewidth,clip,trim=0px 105px 0px 2px]{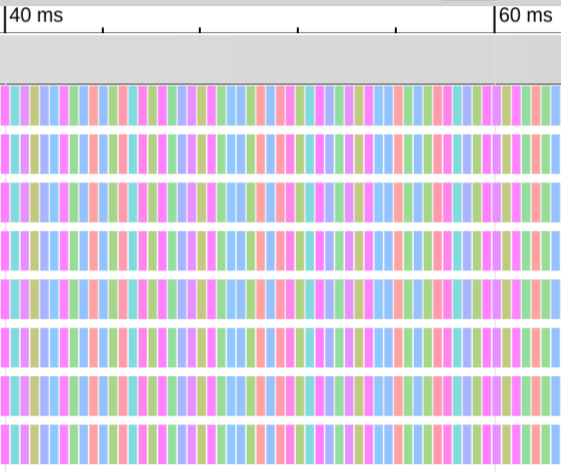}\\
    \subcaption{{\small 16 clients}}
    \label{subfig:bench0concurrent-trace-16-short}
\end{subfigure}
\captionshrink
\caption{Traces of a sample of TPU cores for Figure~\ref{fig:bench0concurrent}. \ourname showing interleaving of gang-scheduled concurrent programs.}
\label{fig:bench0concurrent-trace-short}
\vspace{-1ex}
\end{figure*}
 
\begin{figure*}[th!]
\vspace{-1ex}
\centering
\includegraphics[width=0.83\linewidth]{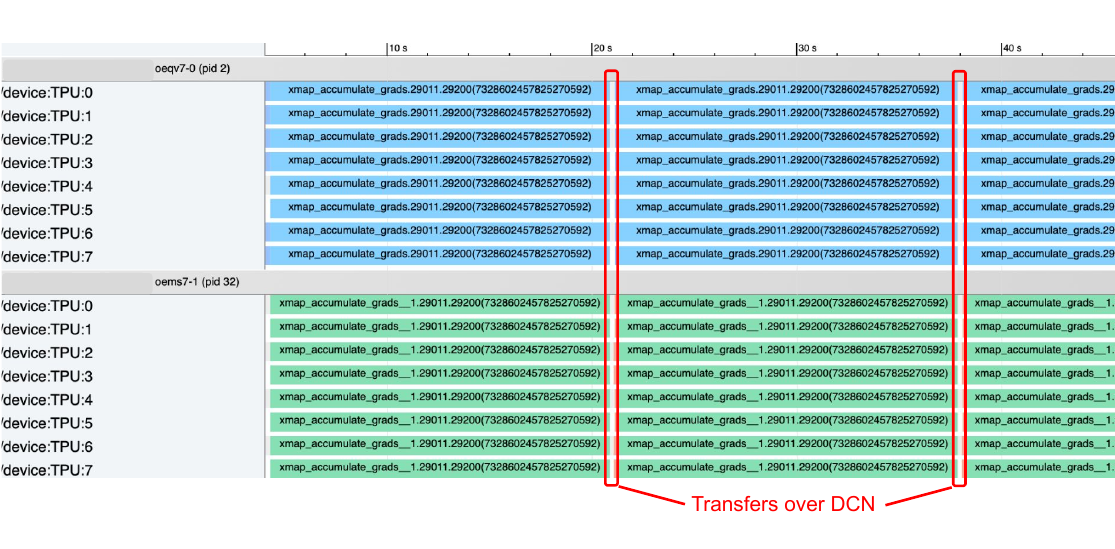}
\captionshrink
\vspace{-4ex}
\caption{64B Transformer model training data parallel over two islands of 512 TPUs each. The trace highlights the relatively small overhead of cross-island transfer using DCN.}
\label{fig:dataparallel-xprof}
\vspace{-2ex}
\end{figure*}

\section{Input Data Processing}
\label{sec:app:input_data_processing}

JAX has deliberately avoided re-implementing data loading pipelines, and \texttt{tensorflow/datasets}~\cite{tfds} are commonly used for JAX input processing, so it is not difficult for JAX programs to be adapted to offload input processing to the CPU-based TensorFlow executors run on \ourname workers.
\ourname instantiates a CPU-based TensorFlow executor on each host, so that user programs can serialize input processing into a TensorFlow graph and distribute it across the workers. We plan to support streaming data protocols so that CPU-based computation can be performed on an independently managed set of servers, thus decoupling the expensive TPU-connected hosts from the CPU resources available for input processing.

\section{Evaluation Workload Traces}\label{sec:app:additional_result}

Figure~\ref{fig:bench0concurrent-trace-short} presents the traces for the workload of Figure~\ref{fig:bench0concurrent} with a varied number of clients submitting programs concurrently (\S\ref{sec:eval-multi-tenancy}).
A single client uses a very small per-program compute time of 0.33~ms that is insufficient to saturate accelerators.
With \ourname's multi-tenency support, using multiple clients increases the device utilization to $\sim{}100$\%.
All client programs are gang-scheduled across all cores, and interleaved at a millisecond scale or less, showing little context-switch overhead.

Figure~\ref{fig:dataparallel-xprof} shows a trace profile for multiple training steps when the 64B Decoder only Transformer model is trained data parallel over two islands of accelerators with 512 chips each (\S\ref{sec:large_scale_models}). The first eight rows (blue) correspond to TPU computations on a host in the first island and the next eight rows (green) correspond to TPU computations on a host in the second island. In this case, each island computes gradients and then enqueues the transfers of these gradients to the other island. When the transfer of gradients is over DCN completes, each island applies the received gradients and starts the next training step. DCN transfers incur minimal overhead even at the scale of pairs of 128 hosts resulting in 97.2\% training throughput compared to an SPMD configuration that uses ICI communication over total equivalent number of chips.

\clearpage

\end{document}